\documentclass[twocolumn,preprint]{aastex62}

\usepackage[T1]{fontenc}
\usepackage[english]{babel}
\usepackage{natbib}
\usepackage{ulem,xcolor}
\usepackage{amsmath, graphicx}
\usepackage{soul}
\usepackage{lineno}
\usepackage{hyperref}

\begin{document}


\title{Investigation of possible precursors of solar flares in the active region 12230 On December 9, 2014}%

\author[0000-0001-7856-084X]{G.G. Motorina}
	\affil{Central  Astronomical Observatory at Pulkovo of Russian Academy of Sciences, St. Petersburg, 196140, Russia}

\affil{Space Research Institute of Russian Academy of Sciences, Moscow, 117997, Russia}

\author[0000-0002-5719-2352]{I.N. Sharykin}
	\affil{Space Research Institute of Russian Academy of Sciences, Moscow, 117997, Russia}

\author[0000-0001-6995-3684]{I.V. Zimovets}
	\affil{Space Research Institute of Russian Academy of Sciences, Moscow, 117997, Russia}

\author{A.S. Motorin}
	\affil{ITMO University, 197101 St. Petersburg, Russia}

\received{February 25, 2025 }

\begin{abstract}
The question of the nature of precursors of solar flares, as well as their relationship with subsequent flares, still has no unambiguous answer. This is due, in particular, to the lack of systematic statistical work, the relative incompleteness (in isolation from the context of the development of the entire active region) of studies of individual events and the ambiguity of the term 'precursor' itself. 
In this paper, we consider the dynamics of the NOAA 12230 active region (AR), in which a series of homologous flares (C5-C9) occurred on December 9, 2014 within 12 hours, with an average frequency of about 2 hours. This AR was characterized by a rapid increase in flare activity followed by a rapid decay, which can be considered as a good example for studying potential precursors of flare series. We investigated the evolution of AR 12230 over a relatively long period (several days) and its transition from the 'no-flare' state to the flare-active regime. For this purpose, we study the magnetic field dynamics using SDO/HMI vector magnetograms, UV-EUV images based on SDO/AIA data, and X-ray observations from GOES/XRS and RHESSI data. Thus, we identified several phases of AR development in terms of magnetic-field dynamics and flare activity. We proposed a method for building hourly integral maps of UV variations from AIA 1600 \AA\ data. We conclude that the significant increase in the chromospheric radiation variations against the background of a small flux of soft X-ray and UV emission from the corona observed on December 8, 2014, together with the magnetic flux emergence, can be considered as a precursor to a series of flares. 
We also analyzed the appearance of X-ray sources of weak bursts before the series of flares. The X-ray bursts developed in the same plasma structures where future flares would occur. The results obtained show the importance and prospects of applying new methods of synoptic observations of the Sun in the context of collecting statistics (“history”) of the AR energy release in different ranges of the electromagnetic spectrum. 
In other words, it is important to monitor not only the dynamics of the magnetic field structure but also how the AR releases the stored magnetic energy. An integrated approach will make it possible to develop new methods of flare prediction: perhaps better than just taking into account the magnetic field structure. 
\end{abstract}



\section{Introduction}
\label{S_Intro}
A precursor to a solar flare (“in traditional meaning”) is usually considered as a localized, small-scale energy release visible as a brightening in various wavelength ranges over a relative time range from a few minutes to ~1-2 h before the main flare (see, e.g., Wang et al., 2017). Precursors can be both a manifestation of independent weaker episodes of energy release accompanying the evolution of the active region (AR) and a trigger for the main flare. In a broader sense, solar flare precursors can be understood as a wide variety of phenomena in the AR that indicate the instability of magnetic-plasma configurations. The search for precursors is closely related to the problem of accumulation and subsequent energy release in ARs during solar flares, as well as to the applied problems of forecasting powerful active phenomena on the Sun, including space weather forecasting. 

For example, Van Hoven and Hurford (1986) described the “classical” flare precursor for a flare and coronal mass ejection (CME) as the appearance of a burst of soft and hard X-ray emission, indicating the preliminary heating of the plasma and acceleration of electrons before the main flare energy release. In addition, a spatial relationship between the small-scale energy release (precursor) near the neutral line (NL) of the magnetic field and the onset of the main flare was shown in (Wang et al., 2017). At the same time, studies of radio and X-ray precursors (Chifor et al., 2007; Abramov-Maximov and Bakunina, 2022; Zimovets et al., 2022; Shohin et al., 2024; Motorina et al., 2023) showed that the sources of preflare fluctuations are usually localized in compact AR zones located near the NL not far (approximately within 20-30 arcseconds) from the sources of the main flare. This suggests that the small-scale energy release in the lower solar atmosphere may be associated with the onset of the main flare. Nevertheless, the question of precursors and their relation to the onset of solar flares remains open.

Notably, not only X-ray and microwave bursts are considered as precursors of solar flares, but also other morphological structures with different spatial scales and temporal dynamics in different ranges of the electromagnetic spectrum. The most prominent manifestations of pre-flare activity are systematized in the list below: \\
\begin{enumerate}
\item Compact brightenings and bursts in different bands of the electromagnetic (EM) spectrum: in the ultraviolet (UV) and extreme ultraviolet (EUV) ranges before solar flares (Chifor et al., 2006; Chifor et al., 2007; Zimovets et al., 2022; Shohin et al., 2024) in the thermal soft X-ray range (e.g., Chifor et al. 2006; Awasthi et al. 2014, 2018a, 2018b; Zhang et al., 2015; Zimovets et al., 2022), gyrosynchrotron bursts of microwave radio emission (Kai et al., 1983; Wang et al., 2017; Bakunina et al., 2020a, 2020b);\\
\item Pre-flare hot channels visible in the EUV range (Cheng et al., 2014; Nindos et al., 2015; Hernandez-Perez et al., 2019; Sharykin et al., 2020);\\
\item Vertical rise of magnetic loops and eruptive magnetic  flux rope prior to the onset of solar flare and corresponding eruption (Ohyama and Shibata, 1997; Zhang et al., 2012; Zhang et al., 2015; Wu et al., 2016; Mitra et al., 2019); \\
\item Long-lived extended sources of emission in different ranges of the EM spectrum: e.g., radio sources above NL (Neutral Line Sources, e.g., see Uralov et al., 2008; Bakunina et al., 2015; Abramov-Maximov et al., 2015) and sigmoids visible in the soft X-ray range (Gibson et al., 2002; Driel-Gesztelyi et al., 2015; Jiang et al., 2014);\\
\item Plasma flows found from UV/ EUV spectroscopy and visual analysis of time-sequence images in the same wavelength range (Wallace et al., 2010, Dudik et al., 2016; Zhou et al., 2016; Woods et al., 2017; Huang et al., 2019);\\
\item Spectroscopy of various UV lines reveals their non-thermal broadenings and Doppler shifts, which may be related to the activation of turbulent flows in pre-flare magnetic structures (e.g., Jeffrey et al., 2018; Huang et al., 2019);\\
\item Quasi-periodic pulsations in pre-flare emission fluxes (e.g., Zhdanov \& Charikov, 1985; Tan et al., 2016; Abramov-Maximov \& Bakunina, 2022; Zimovets et al., 2022) and oscillatory motions in the magnetic structures of pre-flare ARs (Zhou et al., 2016).
\end{enumerate}

All these phenomena have been described in the vicinity (not more than 1-2 hours) of individual solar flares. However, in our opinion, it is of great interest to investigate the set of energy release phenomena in ARs before large flare series, which can have a powerful integral effect on space weather, on long (several days) time scales. What is the character of the behavior of emission sources in different ranges of electromagnetic waves before a series of flares on long time scales of AR dynamics? What is the relationship between the dynamics of pre-flare emission sources and the dynamics of the photospheric magnetic field in the AR?

The purpose of this work is to investigate a specially selected (see the selection criteria in the next section) NOAA 12230 in the time interval of its highest flare activity on December 9, 2014 (as well as about a day before and after) to search for precursors of a series of flares from multiwavelength observations, including soft X-ray (GOES/XRS, White et al., 2005) and RHESSI (Lin et al., 2002), extreme ultraviolet (SDO/AIA, Lemen et al., 2012) emission, photospheric magnetic field data (SDO/HMI, Scherrer et al., 2012). In the present work, we attempt to highlight typical processes in different ranges of the spectrum and in different layers of the solar atmosphere in AR before a series of flares and discuss the possible nature of precursor occurrence. 

In our opinion, the novelty of this work lies in our attempt to conduct a comprehensive multiwavelength (magnetic field analysis + analysis of emission from the chromosphere to the corona) study of the energy release of AR before a series of solar flares on long time scales. We consider the activity “overall” for the selected AR and do not analyze individual events in detail. It is worth noting that most modern works usually estimate the peculiarities of the magnetic field dynamics in comparison with the soft X-ray fluxes based on GOES and do not analyze in detail the statistics of the emission sources during the AR evolution. Here we will gather statistics of emission for a specific, selected AR in different ranges of the EM spectrum. The novelty of this work also lies in the selection and consideration of the excellent case of NOAA 12230 (more details in the next section), which illustrates in a clear way the evolution of the AR from a non-flare to a flare-active state and back from the point of view of the simultaneous analysis of multiwavelength observational data and magnetic field.

\section{SELECTION OF THE ACTIVE REGION FOR THE ANALYSIS}
\label{section2}
For the analysis we selected NOAA 12230, which appeared on the visible part of the Sun on December 6, 2014 (during the maximum of the 24th cycle of solar activity), and in the process of development, beginning on December 8, produced a series of C-class solar flares. This AR was chosen to analyze the dynamics of flare activity in the UV and EUV energy ranges due to the rapid growth of the group of sunspots (within a day) and the appearance of a series of solar flares, while the seven most powerful flares appeared within 12 hours. In our opinion, this AR is a remarkable object to study the properties of the energy release of AR before an isolated series of solar flares. In fact, it is difficult to find cases for large ARs (in which a series of powerful M- and X-class flares are observed) where we see the birth of an AR, followed by an active flare period and a decay of flare activity (see, e.g., Fursyak et al., 2020; Nechaeva et al., 2024). At the same time, this AR also needs to be located in the central part of the solar disk in order to have an acceptable quality of the HMI vector magnetograms. The chosen AR is distinguished by the fact that we see its evolution from the begining to the end of the flare-active period within about three days. At the same time we have good vector magnetograms without the projection effect (characteristic of near-limb ARs). Also, relatively weak C-class flares did not produce strong saturation of the AIA data, which is also convenient for analyzing this AR. Based on the observations of this AR, we will collect unique information on the dynamics of emission sources in different bands of the EM spectrum with respect to the dynamics of the photospheric magnetic field structure at different stages of AR development. In particular, we will emphasize the peculiarities of the AR transition to a series of flares.

\section{OBSERVATIONAL DATA AND PROCESSING}
\label{section3}
The first flare occurred on December 8, 2014 at 17:53 UT (C1.3), then a series of 11 flares occurred consecutively on December 9, 2014 starting at 03:21 UT (C2.0) to 23:22 UT (C1.3), and the last flare occurred on December 10, 2014 at 02:03 UT (C1.4), see Table \ref{table1}. The flares, X-ray class, start and end times are taken from the list freely available at: 
 \url{https://umbra.nascom.nasa.gov/goes/eventlists/goes_event_listings_HER/}. It should be noted that the flare that occurred on December 9, 2014 at 02:33 UT (C1.3) in NOAA 12230 (according to goes\_xray\_event\_list\_2014) was not included in our list because it occurred on the western limb of the Sun and is not related to the considered AR. The duration between the end of one flare and the beginning of the next for December 9, 2014 varied from 51 min. to 3h. 30 min. The mean time between neighboring flares was 100.0 min, the standard deviation was 50.4 min.

\begin{table*}[ht]\centering
\caption{List of solar flares that occurred from December 8 to 10, 2014 in NOAA 12230. From left to right: date, time of the start, peak, end of the flare, X-ray class according to GOES, time between the end of one flare and the start of the next. 
\hspace{1.7in}}
\begin{tabular}{l l l l l l}
\hline\hline
Date & Start time & Peak time & End time & X-Class & Time between flares 
\\ [0.5ex]
\hline
\quad 8-Dec-2014 & 17:53 & 17:56 & 18:00 & C1.3  & -
\\
\quad 9-Dec-2014 & 03:21 & 03:28 & 03:47 & C2.0   	
\\
\quad 9-Dec-2014 & 04:56 & 05:15 & 05:39 & C1.4 & 1 h 9 min
\\
\quad 9-Dec-2014 & 08:22 & 08:30 & 08:41 & C8.1 & 2 h 43 min
\\
\quad 9-Dec-2014 & 09:58 & 10:24 & 10:35 & C8.6 & 1 h 17 min
\\
\quad 9-Dec-2014 & 12:29 & 12:34 & 12:45 & C5.4 & 1 h 54 min
\\
\quad 9-Dec-2014 & 13:43 & 13:48 & 13:56 & C4.5 & 58 min
\\
\quad 9-Dec-2014 & 15:21 & 15:28 & 15:33 & C6.2 & 1 h 25 min
\\
\quad 9-Dec-2014 & 17:12 & 17:18 & 17:23 & C2.5 & 1 h 39 min
\\
\quad 9-Dec-2014 & 18:37 & 18:48 & 18:55 & C5.3 & 1 h 14 min
\\
\quad 9-Dec-2014 & 19:46 & 19:49 & 19:52 & C1.4 & 51 min
\\
\quad 9-Dec-2014 & 23:22 & 23:27 & 23:32 & C1.3 & 3 h 30 min
\\
\quad 10-Dec-2014 & 02:03 & 02:13 & 02:21 & C1.4 & -
\\
[1ex]
\hline
\end{tabular}

\label{table1}
\end{table*}

\begin{figure*}\centering
\includegraphics[width=14cm]{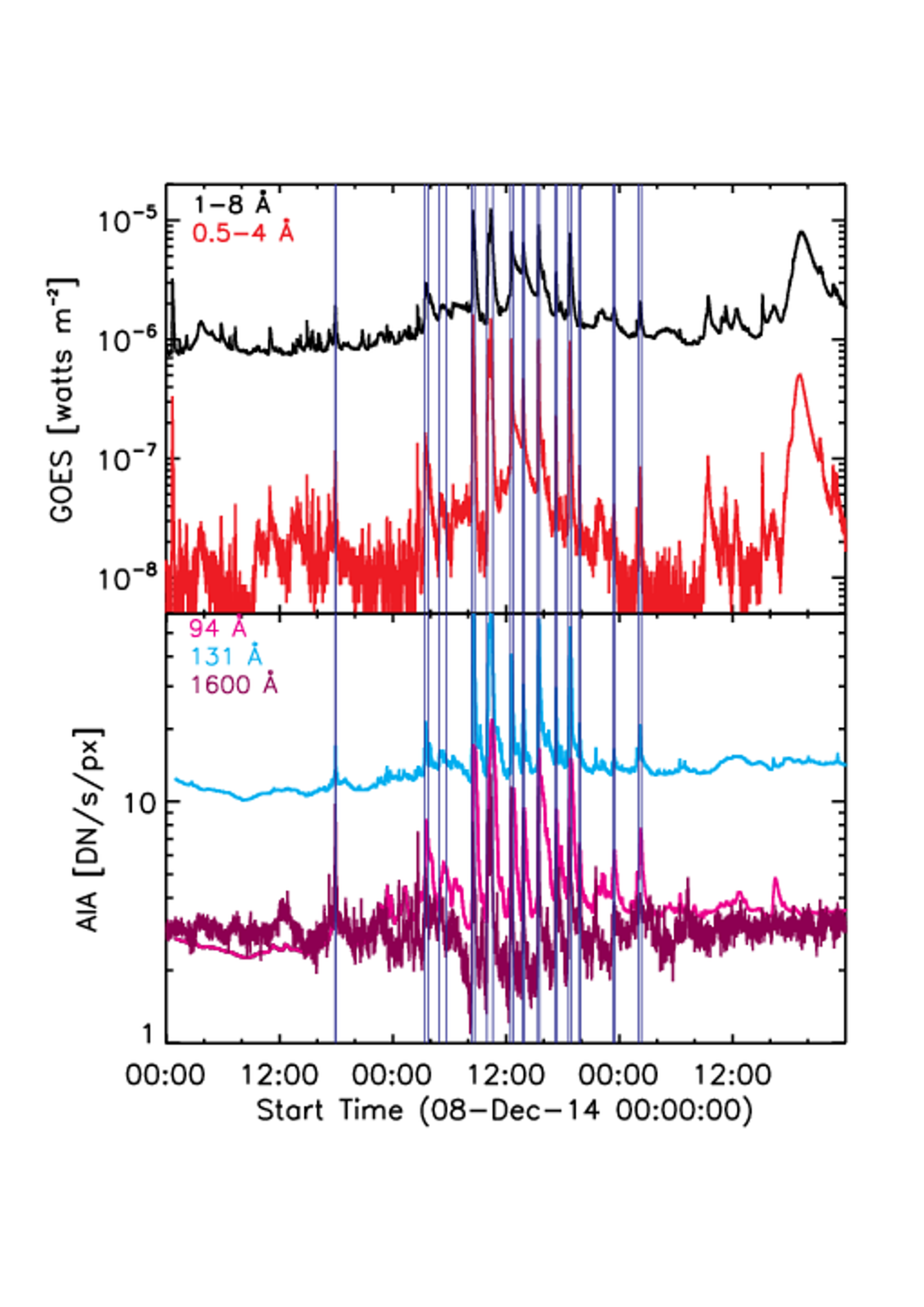}
\caption{The time profiles of the NOAA12230 evolution on December 8-10, 2014 in the GOES X-ray flux in channels 1-8, 0.5-4 \AA\ (upper panel), the EUV flux in SDO/AIA channels 94, 131 \AA\, and the total variation of the UV emission (in relative units) in channel 1600 \AA\ (lower panel). Vertical lines show the start and the end of the flares.}    
\label{Fig1}
\end{figure*}

Figure \ref{Fig1} shows the time profiles of X-ray emission according to GOES data for December 8-10, 2014 in 1-8, 0.5-4 \AA\ channels, EUV emission (94, 131 \AA) from the region shown in Fig.\ref{Fig5}, and the total variation of the UV emission in NOAA 12230 in the 1600 \AA\ channel from the SDO/AIA data. We note the similarity of the temporal profiles of the X-ray and UV emission (Fig.\ref{Fig1}), which consisted of several peaks.

\begin{figure*}\centering
\includegraphics[width=18cm]{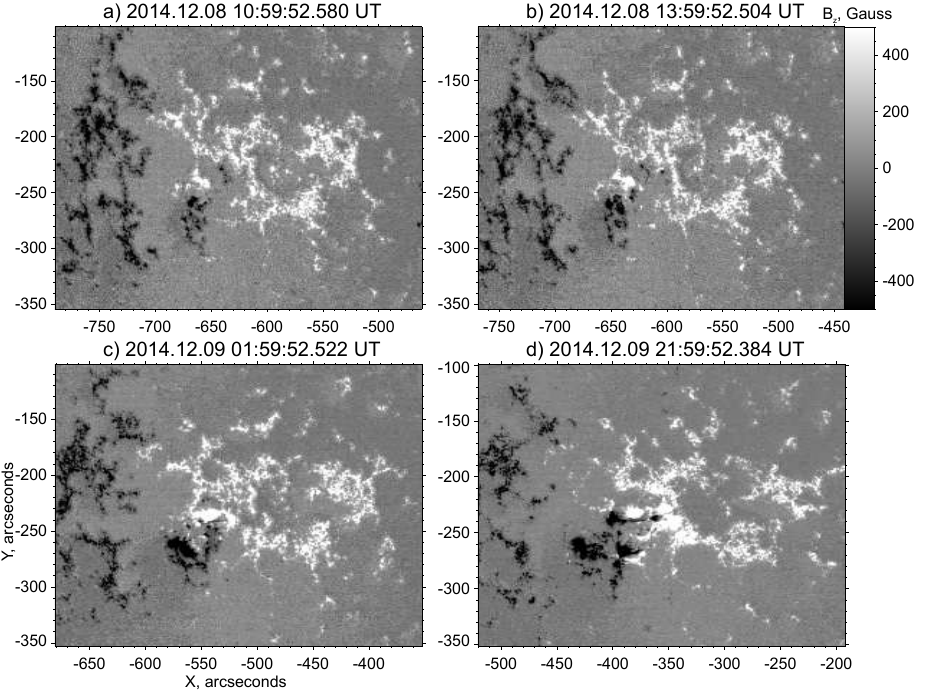}
\caption{The time sequence of SDO/HMI magnetograms for NOAA 12230: Bz component maps at diffetrent times (left to right from top to bottom): 08.12.2014 10:59:52 UT, 08.12.2014 13:59:52 UT, 09.12.2014 01:59:52 UT, 09.12.2014 21:59:52 UT.}    
\label{Fig2}
\end{figure*}

For the analysis of the dynamics of the total vector magnetic field, the SDO/HMI vector magnetograms were examined. Further, we took a fixed area that covered the entire AR for further calculations. Fig.\ref{Fig2} shows the time sequence of the SDO/HMI magnetograms (maps of the Bz component after recalculation into spherical coordinates) for December 8 and 9, 2014. The time range of the rapid formation of NOAA 12230 is clearly distinguished: ephemeral region (only Plage regions are visible) (a), appearance of regions of strong magnetic fields of sunspots (b), relative separation of sunspots and growth of their area (c, d). Vertical fields of more than $\sim$1000 Gauss appeared in the process of the sunspots formation (see below for more details).

\begin{figure*}\centering
\includegraphics[width=14cm]{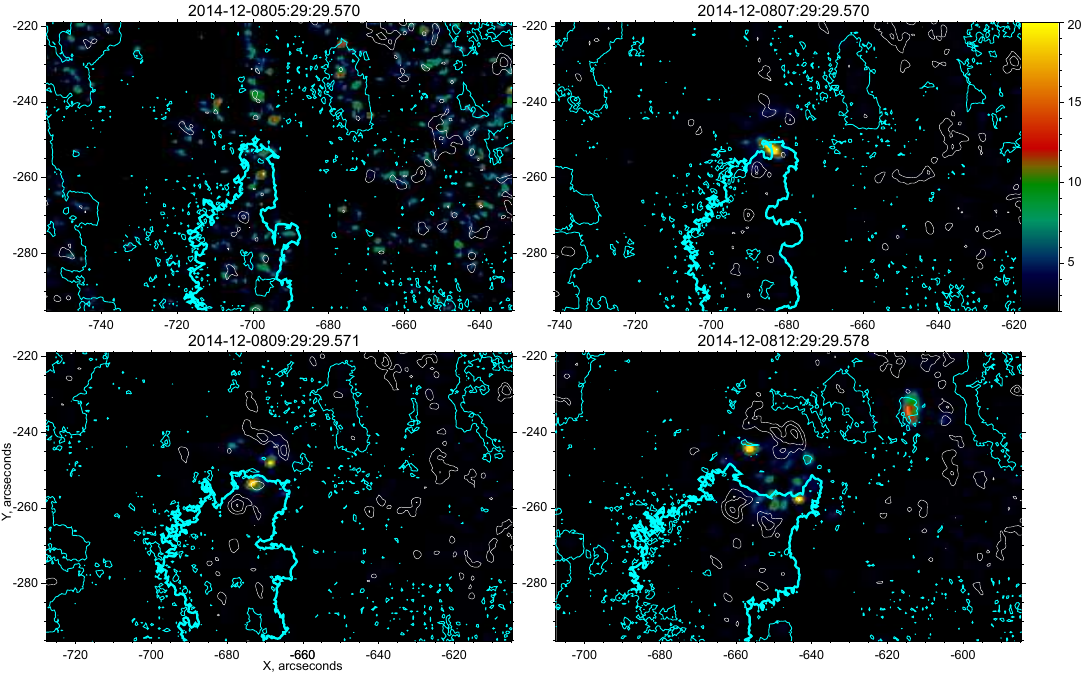}

\includegraphics[width=14cm]{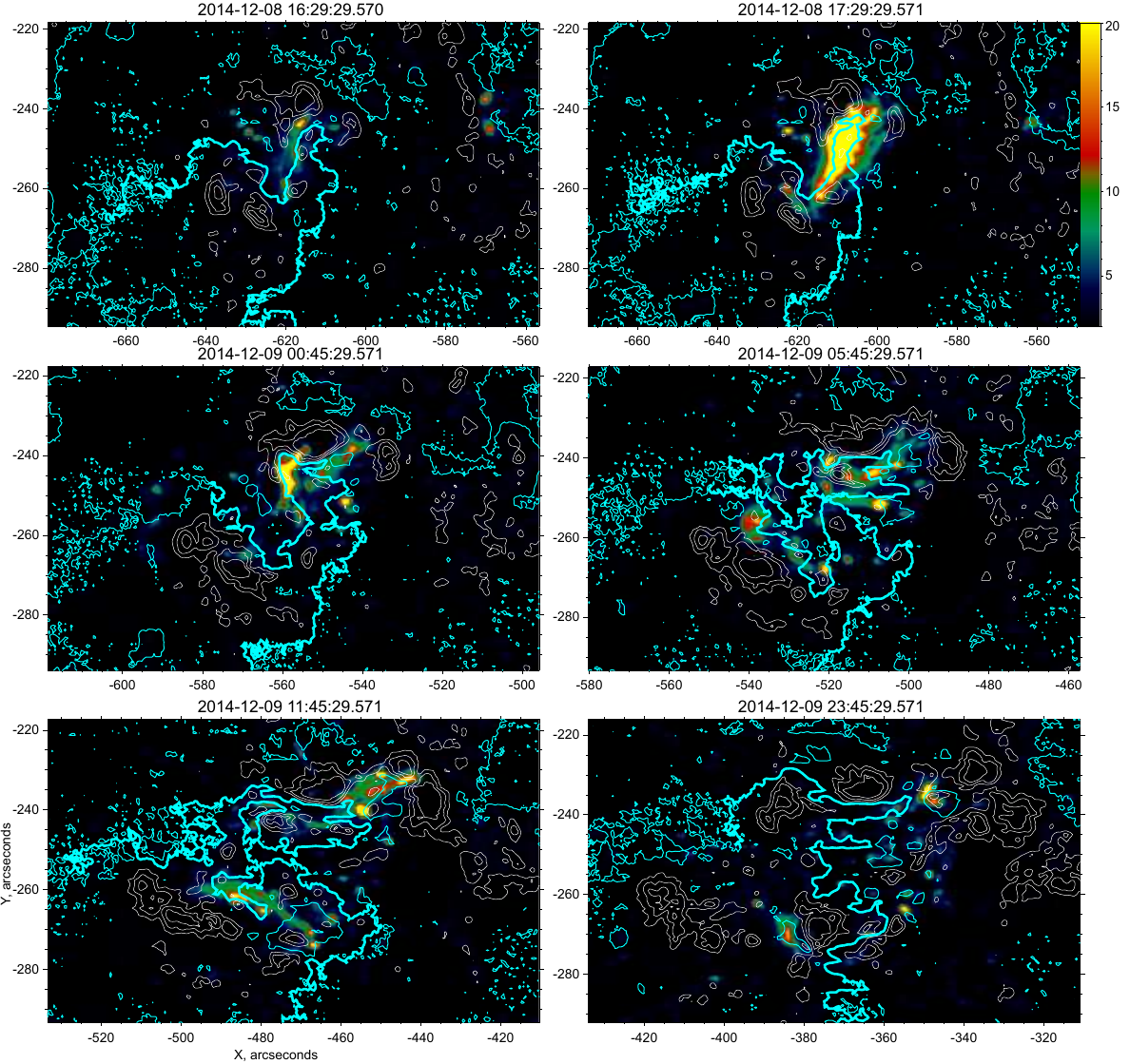}
\caption{The time sequences of total (averaged) UV variance maps of brightenings (colored background) in the SDO/AIA 1600 \AA\ channel. These maps were made by summing in hourly time intervals all positive time derivatives (in pixels). The white contours show the magnetic field modulus levels (500, 1000, and 1500 Gauss). The bold line shows the position of the neutral line.}    
\label{Fig3}
\end{figure*}

Figure \ref{Fig3} shows the hourly integral (i.e., with hourly summation) maps of UV brightenings in the SDO/AIA 1600 \AA\ channel (24 s temporal resolution). The field of view in the figure panels is not fixed and grows as the sunspots appear and the AR grows for better visualization. The total (i.e., integral) UV variance maps of brightenings were constructed from the time sequence of the SDO/AIA UV images as follows. The difference images of NOAA 12230 were made, and then the pixels with negative time derivative were taken to be zero. Pixels in which the brightness growth per frame was less than 150 DNs/frame (DN - digital number) were also set to zero. 
Then at hourly intervals such difference maps (with pixels with significant brightness growth, i.e. more than 150 DNs/frame) were summed up and the value in each pixel was normalized by the number of frames in the selected hourly interval ($\sim$150 difference frames). In this way, maps of the integral hourly ”positive” energy release were generated. Based on this, we compared the dynamics of chromospheric activity in the integral UV maps with the averaged (for 1 hour) vector magnetograms (contours in Fig.\ref{Fig3}): overlaying of the area of strong magnetic fields and drawing of the position of the neutral line on the UV maps.

The four upper panels of Fig.\ref{Fig3} show the appearance of sunspots (flux emergence) and the beginning of AR formation: a bipolar structure appears ($\sim$25 arcsec) and the first compact UV brightenings with a characteristic size up to $\sim$5 arcsec. Most likely, the first brightenings appear as a result of the interaction of the emerging magnetic flux with the background magnetic field. We then observe a complexification of the NL and the appearance of large-scale UV brightening regions: the most powerful flares are shown for 17:29:29 UT. For  times 05:45:29 and 11:45:29 UT, we see a strongly fractured NL and many UV brightenings in the vicinity of this NL. The bottom panel of Fig.\ref{Fig3} shows the time frame when we no longer see increased solar activity and the UV brightenings again become quite sparse and compact. 

\begin{figure*}\centering
\includegraphics[width=18cm]{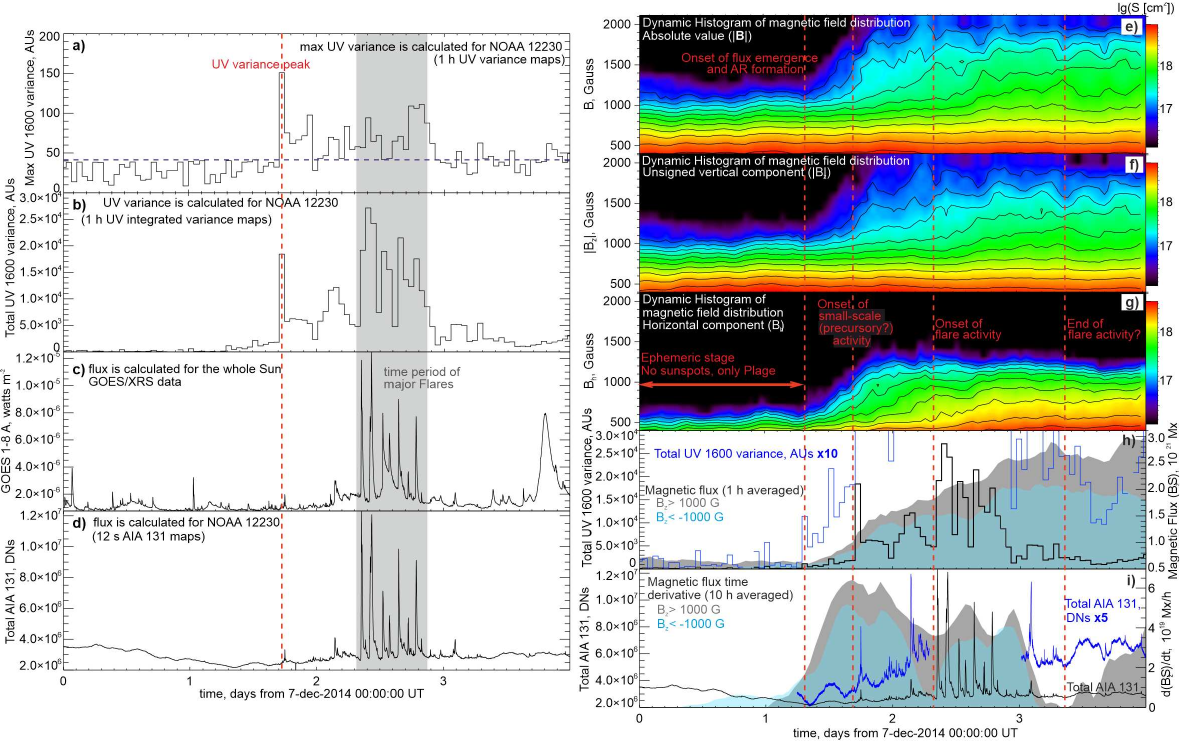}
\caption{Time profiles: the maximum level of variations in UV emission (brightest pixel in the UV maps) in NOAA 12230 in the 1600 \AA\ channel (a), the total variation in UV emission in NOAA 12230 in the 1600 \AA\ channel (b), the GOES X-ray flux in the 1-8 \AA\ channel (c), the EUV flux in the 131 \AA\ channel (d), histograms of distribution of magnetic field components ( the X-axis shows time and the Y-axis shows the histogram): field modulus (e), modulus Bz component (f), horizontal B component (g). The color scale of the histograms shows the logarithm of the total area of the regions (in cm$^2$) for a particular value of the magnetic field component (marked on the vertical axis). Vertical lines indicate characteristic stages in the development of NOAA 12230 and background energy release. The panels (h, i) show amplitude-increased time profiles of the UV emission in the 1600 \AA\ channel and EUV emission in the 131 \AA\ channel (blue) for better visualization of weak bursts. }    
\label{Fig4}
\end{figure*}

\begin{figure*}\centering
\includegraphics[width=18cm]{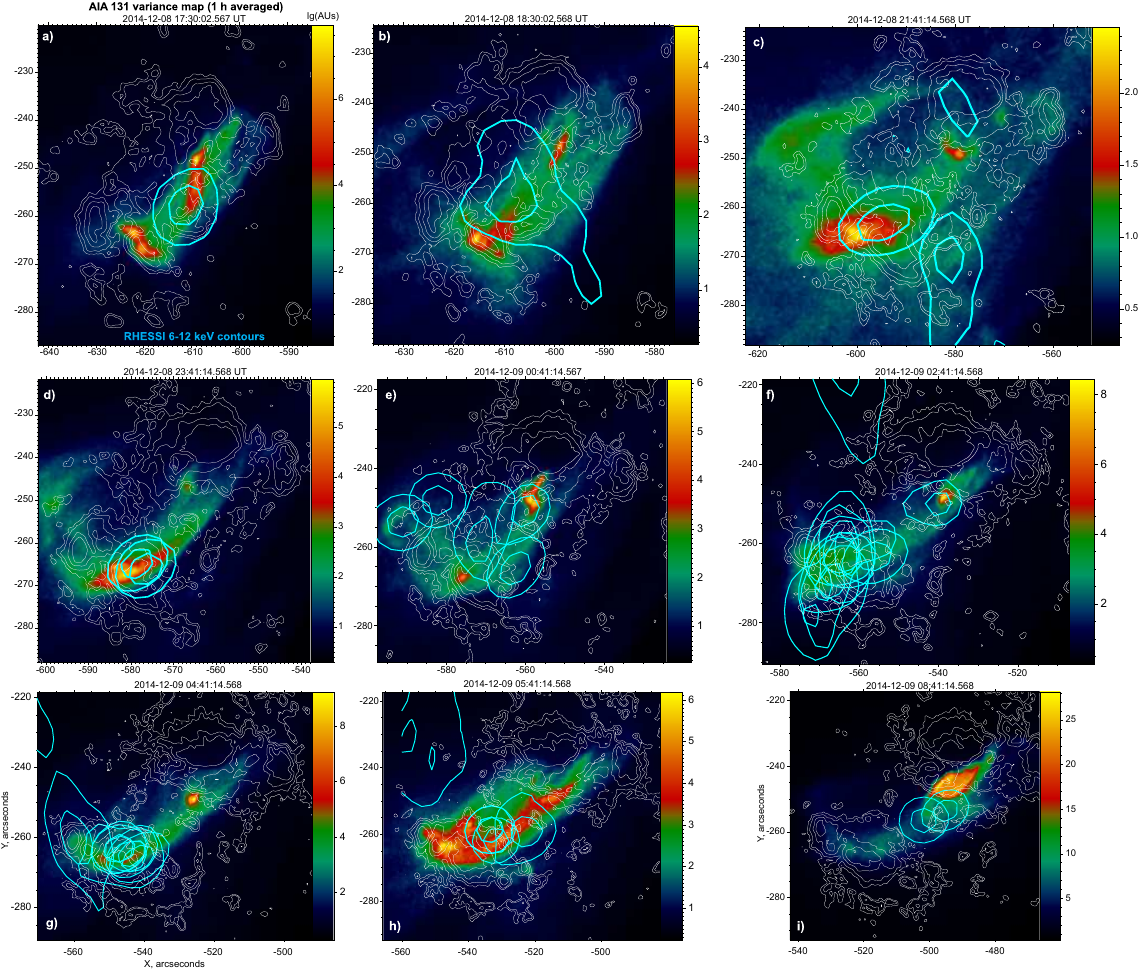} 
\caption{The RHESSI images of 6-12 keV X-ray sources (detectors 3, 6, 7). Time sequences of the EUV brightenings maps (colored background) in the 131 \AA\ channel. White contours show the magnetic field modulus levels (500, 1000, 1500, and 2000 Gauss). Cyan contours show the positions of the sources in the 6-12 keV energy range.}    
\label{Fig5}
\end{figure*}

Fig.\ref{Fig4} shows different temporal profiles of the UV (1600 \AA), EUV (131 \AA), X-ray emission, and histograms of the distribution of magnetic field vector components from the SDO/HMI data. Panels (a) and (b) show the dynamics of the maximum (a, brightest map pixel) and total variation of the SDO/AIA 1600 \AA\ UV emission at hourly summation (from the maps in Fig.\ref{Fig4}). The total EUV 131 \AA\ emission flux from the studied AR is shown in panel (d) in comparison with the integrated flux from the Sun in the soft X-ray range (Fig.\ref{Fig4}c) according to GOES (channel 1-8 \AA). When comparing the UV and X-ray light curves, it can be seen that the GOES data show flares not only from the considered AR, but also from other regions of the Sun, which is important to keep in mind when studying flares and their precursors (Zimovets et al., 2022; Zimovets et al., 2023).

The dynamics of the histograms of the distribution of magnetic field components from the SDO/HMI vector magnetograms (time resolution of 12 min) is shown in Fig.\ref{Fig4}e-g: the modulus of the magnetic field vector (e), the modulus of the vertical (or radial, i.e., perpendicular to the Sun's surface at a given point) component (f), and the magnitude of the horizontal (or tangential, i.e., directed along the Sun's surface) component (g). The phase of magnetic flux emergence and the formation of AR with a group of sunspots from the ephemeral AR stage are clearly distinguished. The color scale shows the total area of pixels with one or another field value (vertical scale) and it can be seen by the purple color that magnetic fields greater than $\sim$700 Gauss (horizontal component) and $\sim$1300 Gauss (vertical component) appear during the magnetic flux emergence process. The lower panels (h) and (i) show the magnetic fluxes computed for the vertical component greater than 1000 Gauss (h: negative and positive) and the time derivatives of these magnetic fluxes (i). These time profiles are compared with the total UV activity (b) and the total EUV 131 \AA\ emission flux for the studied AR (d).

The observational data in Fig.\ref{Fig4} demonstrate the presence of several phases of development of the AR flare activity. In the first ephemeral phase, we observe only the diffuse field of the flare plage and NL, in the vicinity of which there is a local magnetic flux emergence, and the formation of the first sunspots. 
The beginning of emergence corresponds to the first UV brightenings at the chromospheric level according to the AIA data: panel (h) shows in blue the histogram magnified by a factor of 10 in amplitude in order to show the weak increase (in the area of the first vertical dashed line). The maximum value of the magnetic flux variation for both signs of the vertical magnetic field component (second vertical line) is reached at the onset of the small-scale flare coronal activity visible in the 131 \AA\ AIA hot channel (as small bursts compared to flares): the time profile in panel (i) with amplitude magnified by a factor of 5 is also shown for clarity. 
We also see the maximum value of UV variations (based on hourly integral maps) by the curve in Fig.\ref{Fig4}a and the beginning of strong integral chromospheric activity in the AR (Fig.\ref{Fig4}h). The period of small-scale burst activity ends with the beginning of a series of flares (between the third and fourth vertical line), after which the chromospheric activity decays and the number of bursts in the 131 \AA\ channel decreases (after the fourth vertical line).

The obtained analysis of the total maps of UV brightenings and magnetic field data indicates that these maps are sensitive to the magnetic field dynamics. The appearance of brightenings is associated with magnetic flux emergence and the complexity of the magnetic field geometry (visible as NL complexity and sunspot group growth). It is likely that current layers forming in the corona above the AR and the energy release in them lead to energy flows from the corona to the chromosphere, and we see dynamic bright regions in the 1600 \AA\ channel. 
The formation of current layers in the chromosphere is also possible in the case of low magnetic loops. Moreover, we note an important fact that after reaching the maximum change rate of the magnetic flux, we start to see very strong chromospheric activity and, after about 10 hours, the beginning of a series of flares whose chromospheric activity appeared to be comparable to the pre-flare period (with much stronger activity in the corona observed in the X-ray range).

For the analysis of X-ray sources, we constructed images from RHESSI data (Lin et al., 2002) in the 6-12 keV energy range (CLEAN algorithm, Hurford et al., 2002) for all visible bursts in a given AR prior to a series of C-class flares on December 9, 2014. A comparison of the resulting X-ray contour images (Fig.\ref{Fig5}) with the total hourly EUV brightenings maps of 131 \AA\ was made (similar to the UV brightenings maps in Fig.\ref{Fig3}). 
For a given channel, approximately 300 difference frames are within the hourly time interval. In contrast to the total UV variance maps in the 1600 \AA\ channel (predominantly chromospheric emission), the built maps in the 131 \AA\ channel represent the integral hourly activity in the corona and should indicate the most heated magnetic-plasma configurations. When building the total 131 \AA\ variance maps, we summed the pixels of the difference images only for magnitudes of change over 300 DNs/frame (other pixels were set to zero).  We compare the total EUV variance maps and X-ray sources with the full magnetic field contours and do not show the NL (to reduce the clutter in the figure; the NL and the vertical component of the magnetic field see in Fig.\ref{Fig3}).

Groups of X-ray sources were plotted for each individual burst within the considered hourly time interval of the  total EUV variance maps (the beginning of the hourly interval is shown in the captions above the panels in Fig.\ref{Fig5}). Each individual X-ray source was built for sufficiently arbitrary time intervals so as to cover individual bursts or their groups in the RHESSI count rate time profiles in the 6-12 keV energy range, and also so that it was possible to synthesize an X-ray image from the data (to ensure convergence of the image reconstruction algorithm), given the rather poor statistics of the registered X-ray photons. 

Figure \ref{Fig5} shows that X-ray sources and areas of increased EUV activity (bright green-red-yellow areas on the total EUV variance maps) are localized in a single region. At different hourly time intervals, the X-ray sources differ in size and magnitude in relative spread. The first “significant” C-class flare of the series is shown in panel (i). Panel (a) shows the hourly time interval in the vicinity of the maximum pre-flare burst of UV activity (when the third phase of pre-flare microburst activity began after the flux emergence phase). We see that an X-ray source emerges between the paired EUV brightening regions. A weak microburst $\sim$B1.0 was detected for this source, if we subtract the background before this burst, which was at the level of $\sim$C1.0. 

\begin{figure*}\centering
\includegraphics[width=16cm]{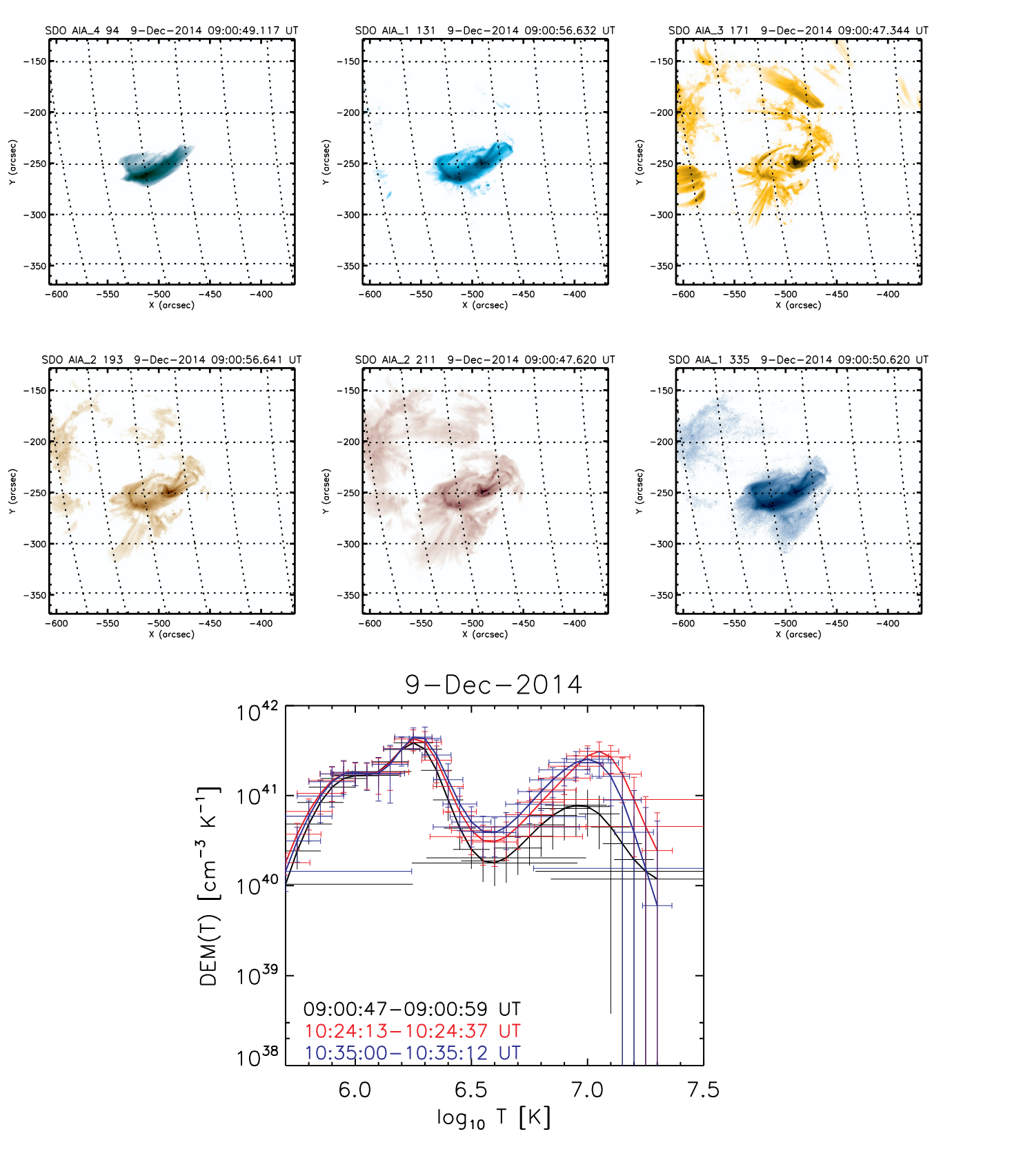}
\caption{Top panel: the analyzed region (FOV) of NOAA 12230, 09:00:49 UT 09.12.2014. Bottom panel: DEM derived from the FOV for three different times: 57 min before the flare SOL20141209T09:58 (C8.6) (09:00:47-09:00:59 UT), at the maximum of the impulsive phase (10:24:13-10:24:37 UT), and during the decay phase (10:35:00-10:35:12 UT).}    
\label{Fig6}
\end{figure*}

The analysis of the UV and EUV emission fluxes does not provide direct estimations of the plasma temperature in the AR. At the same time, one of the possible prognostic parameters of the flare activity may be the average temperature in the AR, or its value in some local places of the AR: for example, the temperature increase may indicate an increasing energy release in the pre-flare current layers distributed over the AR. Within the framework of this work, we will perform a preliminary and simplest analysis of the thermodynamic parameters of the plasma in AR in channels 94, 131, 171, 193, 211, 335 \AA\ from SDO/AIA data and X-ray emission from GOES data. Using the AIA data, we recovered the differential emission measure (DEM) of the pre-flare and flare plasma in the temperature range T = 0.5-25 MK to determine the line-of-sight averaged temperature ($T$) and emission measure ($EM$) in the FOV (field of view, Fig.\ref{Fig6}, bottom panel) of the AR region of interest.

\begin{figure*}\centering
\includegraphics[width=8cm]{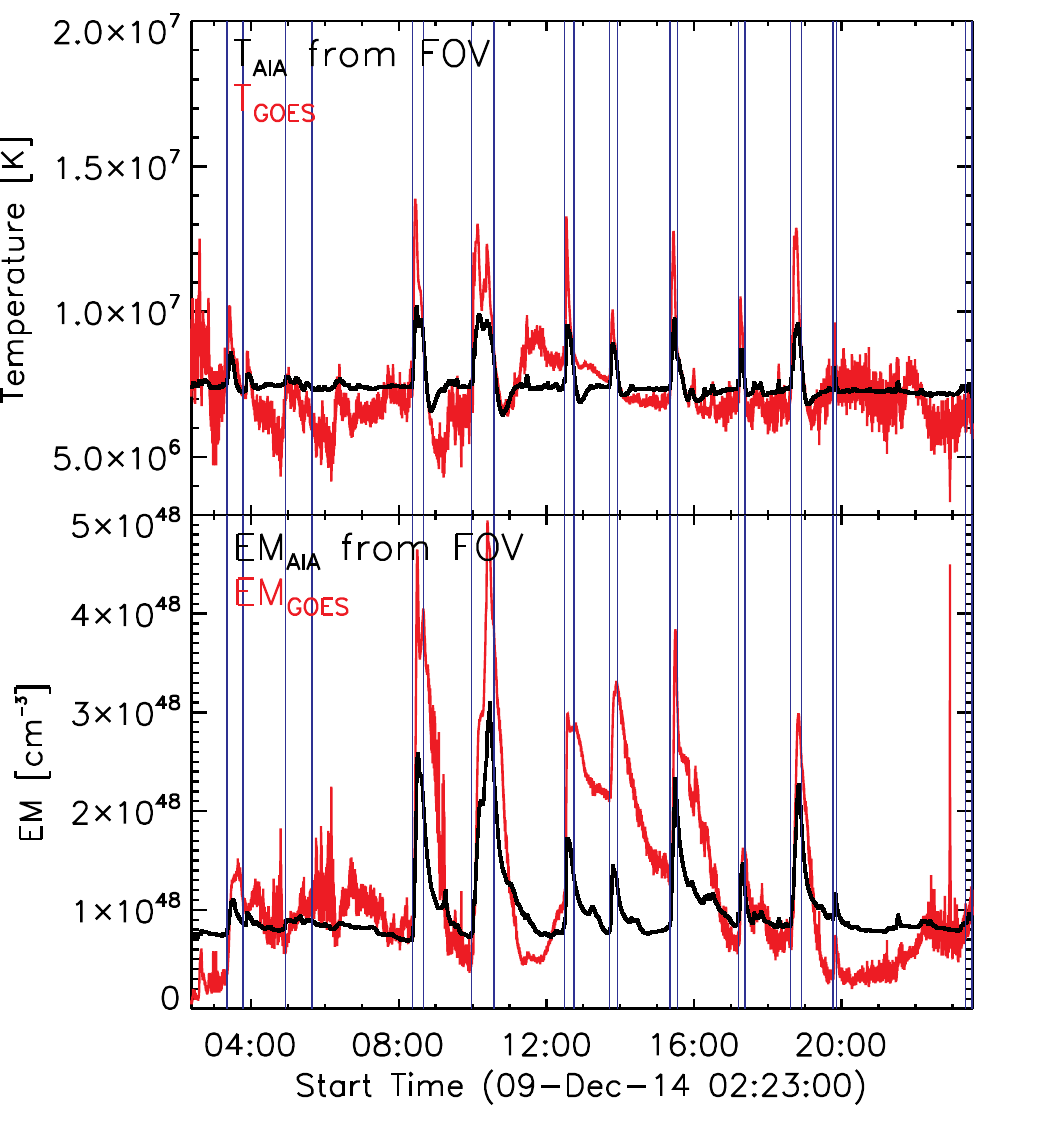}
\includegraphics[width=8cm]{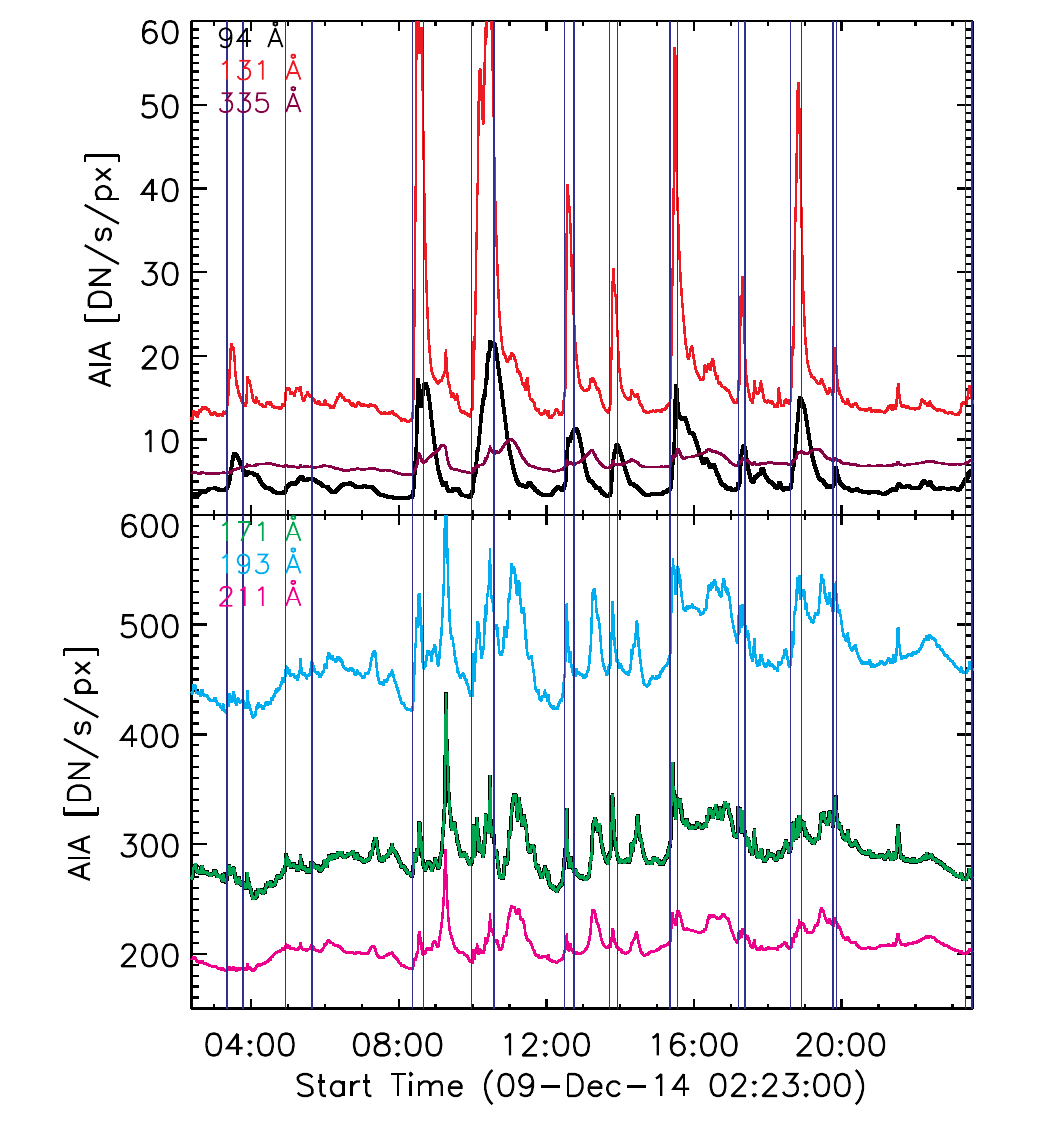} 
\caption{Left: temporal evolution of temperature (T) and emission measure (EM) derived from the GOES and the SDO/AIA data from the FOV shown in Fig.\ref{Fig6} (left panel). Right: light curves of EUV emission from the FOV, December 9, 2014. The vertical lines show the start and end of the flares.}    
\label{Fig7}
\end{figure*}

Using the AIA data with the Tikhonov regularization method (Tikhonov and Arsenin, 1979; Hannah and Kontar, 2012), we performed a reconstruction of the differential emission measure (Fig.\ref{Fig6}, bottom panel). Fig. 6 (bottom panel) shows the DEM calculated from the FOV (Fig. \ref{Fig6}, top panel) for the most powerful flare SOL20141209T09:58 (C8.6) in the flare series for the time moments: 57 min before onset (09:00:47-09:00:59 UT), at the maximum (10:24:13-10:24:37 UT), and during the decline phase (10:35:00-10:35:12 UT). There is DEM increase for coronal plasma with temperatures above 5 MK (as the event progresses). This is most noticeable at the moment of the impulsive phase maximum (10:24:13-10:24:37 UT). The flare-associated heating and growth of the emission measure can be seen in Fig.\ref{Fig7}.

The obtained temporal profiles of temperature and emission measure from GOES and AIA data for a long ($\sim$20 h) time interval are shown in Fig.\ref{Fig7}. It can be seen that the $EM$ and $T$ profiles are correlated. Moreover, $EM$ and $T$ do not show such a clear variation as the temporal profiles of the EUV emission, which can be explained by the fact that $EM$ and $T$ are integral features (averaged along the line-of-sight). It can also be noted that the variations in the $EM$ and $T$ amplitude from AIA are weaker than the variations of $EM$ and $T$ from GOES due to the fact that the analysis of the AIA data was carried out for a selected area of the Sun (GOES measures the integral X-ray fluxes from the Sun as a star), and also AIA is sensitive to lower temperatures of the emitting plasma compared to GOES (e.g., Tsap \& Motorina, 2017).

\section{DISCUSSION OF RESULTS AND CONCLUSIONS}

We analyzed the time sequence of UV and EUV images (chromospheric and coronal emission) from 1600 and 131 \AA\ AIA data of the active region NOAA 12230 before and during the flare activity on December 8-9, 2014, in order to identify the features of the active region's preparation for the flare series of 11 X-ray C-class events with an average time between them of 100 $\pm$ 50 min. A simple efficient method (integrated hourly UV and EUV maps of the positive intensity derivative) for visualizing the chromospheric and coronal brightenings was proposed and demonstrated for a studied AR which best shows the sunspot group formation, the transition to the flare series, and the flare activity decay.

It is shown that the method of integrated UV maps showing chromospheric brightenings is very sensitive to changes (magnetic flux change and NL complication) of the photospheric magnetic field. It is likely that this method of mapping the energy release variations in the AR at the chromospheric level can be used as an additional source of information about the AR energy release status to create new solar flare prediction models (along with information on the photospheric magnetic field). However, extensive statistical work is needed to create such a method. In this paper, we have only demonstrated an example of simultaneous qualitative analysis of data on the photospheric magnetic field, integrated maps of UV chromospheric activity, and the integral flux of coronal EUV emission for a specific selected AR during a few days.

The main physical conclusion of this work is connected with the found phases of the AR preparation for a series of flares. The possibility of obtaining this result is due to the fact that we chose an ideally suited fast evolving AR for the study and different types of data showing different layers of the solar atmosphere. In particular, the method of total UV chromospheric maps allowed us to identify a phase of enhanced chromospheric activity before a series of flares. As a result, five phases of AR dynamics were identified for the studied AR from its birth to the flare activity decay. The boundaries between the phases and their names are rather conventional:
\begin{enumerate}
\item Phase 1. ``Ephemeral AR’’ without sunspots;
\item Phase 2 (``Photospheric’’ phase, AR birth). Magnetic flux started to emerge. Formation of pores and small sunspots. First UV brightenings appear in very compact areas. In this phase, the fluxes of the EUV emission in the 131 \AA\ channel gradually change and microflare bursts are not yet observed;  
\item Phase 3 (``Chromospheric phase’’: small bursts of UV emission). Beginning of microflare burst coronal activity: probably precursors of future flare activity. This phase is observed from a very bright bursts at 1600 \AA\, but with a weak manifestation in the corona (probably very low magnetic loops are involved). The UV brightenings become very bright and cover a large area. Bursts in the 131 \AA\ hot AIA channel begin to be observed. Here we observe the first X-ray microflare burst at ~B1.0 (taking into account the subtracted GOES background). This phase is associated with the formation of a complex NL, probably related to the secondary emergence of the compact magnetic flux against the background of the forming AR;
\item Phase 4 (“Coronal Flare Phase”). Flare activity in the form of a continuous series of C flares. A complex morphology of radiation sources is observed. The cause of the flares is clearly related to the complex NL and photospheric magnetic field structure;
\item Phase 5 (“Flare Activity Decay” in all ranges of the EM spectrum). Decay of flare activity and exit of the AR to the stationary stage. The AR then moves through the disk as an AR that has a Hale class (or Mount Wilson class) of $\beta\gamma$ (see e.g. Toriumi \& Wang, 2019).
\end{enumerate}

We consider X-ray microflare bursts and weak flares (B to C class) in the context of the development of AR NOAA 12230 from its very origin to the onset of flare activity. The complex of observational multi-wavelength data suggests that the location of X-ray sources before the C-class flares is associated with a heated magneto-plasma structure (visible in the integrated EUV maps) in which the energy release of future flares of this AR will develop. Despite the apparent obviousness of this statement, we note that, from our point of view, there has been little detailed analysis of the dynamics of X-ray flare sources (of different powers) over relatively long time intervals. Usually, studies of the solar X-ray emission are tied to specific (more often large over M1.0 class) events. In this work, we have shown by comparing the total maps of EUV variations, HMI vector magnetograms, and X-ray images that the X-ray activity in the AR is localized in the corona above the NL between the spots where the energy release of a series of flares occurs. We note that the importance of processes over NL before flares has been noted in numerous works (some of them are mentioned in Section \ref{S_Intro}): mainly the analysis of individual events and time moments near the onset of flares. In this paper, we have strengthened the statement of the importance of diagnosing high-energy processes in the AR over NL by analyzing processes in the AR over long time intervals. This simple result is important from the point of view of solar flare prediction practice: collecting information on the real status of the AR energy release in the X-ray range (as well as in other ranges of the EM spectrum) in the context of the magnetic field dynamics revealing by photospheric vector magnetograms. In other words, it is important to take into account not only the ``AR energy content’’ (the value of free magnetic energy), but also investigate how the AR releases magnetic energy in the form of flares and small bursts of different energy scales.

In addition to analyzing EUV-UV variance maps and images, it seems important to construct averaged characteristics of the EUV emission by DEM analysis because of the importance of tracking the current status of the AR energy emission. In this work, we analyze the differential emission measure and calculate the averaged (through DEM) flare plasma parameters (T, EM). It is shown that there is a good correlation between the emission measure and temperature profiles obtained from GOES and AIA data, indicating a single process observed with both instruments. A comparison of the AIA coronal channels and GOES data confirmed the presence of bursts from another AR in the X-ray data. The DEM method applied to analysis of AR energy release (over a few or many days) needs to be further developed in the context of developing flare forecasting methods. Here, we investigated the simplest parameters from the integral DEM AR (for the used FOV) with the aim of further developing flare prognostic criteria from X-ray and EUV data.

Consideration of precursors 1-2 h before the flare onset very strongly limits the real possibilities of predicting the energy release of ARs considering slow magnetic field dynamics. This work has shown that it is necessary to consider long time periods of the AR development and also to observe the integral energy release in ARs. In particular, before the beginning of the flare series on December 9, 2014, a significant enhancement of the chromospheric UV activity was seen in the studied AR in about 12 hours (Fig.\ref{Fig4}), which can be considered as a precursor of the flare series against the background of the magnetic flux emergence. In other words, the AR magnetic energy pumping followed first, and then the formation of unstable magnetic configurations and, in particular, current layers occurred. With this approach to data analysis, the concept of a precursor (as a concept for practical flare prediction problems) in a few minutes or tens of minutes loses its practical importance. However, these statements require additional analysis of the observations. Extensive statistical work must be done to realize in practice the flare forecast method based on multiwavelength observations of AR energy release.

It is worth noting that almost past and present studies of microflares have been conducted in the context of studies of coronal heating, or as detailed multi-wavelength studies of individual events. In this work, it has been shown that the energy release in the form of relatively weak bursts in the developing AR is related to the future energy release of solar flares. It seems that small acts of energy release are indicative of increasing instability and future flares.

We emphasize the main methodological conclusion of this work. We have demonstrated a qualitative approach to the analysis of multi-wavelength observational data of a particular flaring AR and have shown the possibility of collecting (by constructing total UV/EUV maps and DEM analysis) information on activity of different energy scales in the context of the photospheric magnetic field dynamics. Further it is necessary to proceed to statistical studies, search for new methods and development of quantitative parameters indicating the current and previous (relative to a given time moment) status of the AR energy release in order to predict future energy release using also time series of photospheric vector magnetic field data. From our point of view, progress in empirical models of solar activity forecasting will be related to the addition of multiwavelength data (showing different layers of the solar atmosphere) with a sufficiently good spatial resolution to the usual used vector magnetograms.

\section*{ACKNOWLEDGMENTS}
The work was supported by Russian Science Foundation grant no 20-72-10158.


\section*{bibliography}



Abramov-Maximov V.E., Borovik V.N., Opeikina L.V., et al. // Sol. Phys. 290 (1), 53 (2015).

Abramov-Maximov V.E., Bakunina I.A. // Ge\& Ae. V. 62. No. 7. P. 895–902. 2022.

Awasthi A.K., Jain R., Gadhiya P.D., et al. // MNRAS. 437 (3). 2249. 2014.

Awasthi A.K., Liu R., Wang H., et al. // Astrophys. J. 857 (2). 124. 2018a. 

Awasthi A.K., Rudawy P., Falewicz R., et al. // Astrophys. J. 858 (2). 98. 2018b.

Bakunina I.A., Melnikov V.F., Morgachev A.S. // Astrophysics 63 (2). 252. 2020a.

Bakunina I.A., Melnikov V.F., Morgachev A.S. // Ge\& Ae. 60 (7). 853. 2020b.

Bakunina I.A., Melnikov V.F., Solov’ev A.A., et al. // Sol. Phys. 290 (1). 37. 2015.

Cheng X., Ding M.D., Zhang J., et al. // Astrophys. J.789 (2). L35. 2014.

Chifor C., Mason H.E., Tripathi D., et al. // Astron. and Astrophys. 458 (3). 965. 2006.

Chifor C., Tripathi D., Mason H.E., et al. // Astron. and Astrophys. 472 (3). 967. 2007.

Dudık J., Polito V., Janvier M., et al. // Astrophys. J. 823 (1). 41. 2016.

Fursyak Yu.A., Abramenko V.I., Kutsenko A.S. // Astrophysics. V. 63. No. 2. P. 260-273. 2020.
 
Gibson S.E., Fletcher L., Del Zanna G., et al. // Astrophys. J. 574 (2). 1021. 2002.

Hannah I.G., Kontar E.P. // A\& A. V. 539. Id. A146. P. 14. 2012.

 Hernandez-Perez A., Su Y., Thalmann J., et al. // Astrophys. J.887 (2). L28. 2019.
 
Huang N., Xu Y., Sadykov V.M., et al. // Astrophys. J.878 (1). L15. 2019.

Hurford, G.J., Schmahl, E.J., Schwartz, R.A., et al. // Sol. Phys. 210. 61. 2002.

Jeffrey N.L.S., Fletcher L., Labrosse N., et al. // Science Advances 4 (12). 2794. 2018.

Jiang C., Wu S.T., Feng X., et al. // Astrophys. J. 780 (1). 55. 2014.

Kai K., Nakajima H., and Kosugi T. // Publ. Astron. Soc. Japan 35 (2). 285. 1983.

Lemen J.R., Title A.M., Akin D.J. et al. // Sol. Phys. V. 275. I. 1-2. P. 17-40. 2012.

Lin, R.P., Dennis, B.R., Hurford, G.J., et al. // Sol. Phys. V. 210. I. 1. P. 3–32. 2002.

Mitra P.K. and Joshi B. // Astrophys. J. 884 (1). 46. 2019.

Motorina, G. G., Tsap, Yu.T., Smirnova, V.V., et al. // Ge\& Ae. V. 63. I. 8. p.1218-1223. 2023.

Nindos A., Patsourakos S., Vourlidas A., et al. // Astrophys. J. 808 (2). 117. 2015.

Ohyama M. and Shibata K. // Publ. Astron. Soc. Japan 49. 249. 1997.

Scherrer, P.H., Schou, J., Bush, R.I., et al. // Sol. Phys. 275. 207. 2012.

Sharykin I.N., Zimovets I.V., and Myshyakov I.I. // Astrophys. J. 893 (2). 159. 2020.

Shohin, T.D., Charikov, Yu.E., Shabalin, A.N. // Ge\& Ae. V. 64. I. 8. p. 1386-1394. 2024.

Tan B., Yu Z., Huang J., et al. // Astrophys. J. 833 (2). 206. 2016. 

Tsap Yu.T., Motorina G.G. // Ge\&Ae. V. 57. I. 7. 2017.

Toriumi S., Wang H. // Living Rev Sol Phys. 16. 3. 2019.

Uralov A.M., Grechnev V.V., Rudenko G.V., et al. // Sol. Phys. 249 (2). 315. 2008.

Van Driel-Gesztelyi L. and Green L.M. // Living Reviews in Solar Physics 12 (1). 1. 2015.

Van Hoven G., Hurford G.J. // ASR, 1986, V. 6, I. 6, pp. 83-91.

Wallace A.J., Harra L.K., Van Driel-Gesztelyi L. // et al., Sol. Phys. 267 (2). 361. 2010.

Wang H., Liu C., Ahn K., et al. // Nature Astronomy 1, 0085. 2017.

White S.M., Thomas R.J., Schwartz R.A. // Sol. Phys. V. 227. I. 2. P. 231-248. 2005.

Woods M.M., Harra L.K., Matthews S.A., et al. // Sol. Phys. 292 (2). 38. 2017.

Wu Z., Chen Y., Huang G., et al. // Astrophys. J.820 (2). L29. 2016.

Zhang J., Cheng X., and Ding M.-D. // Nature Communications. 3. 747. 2012.

Zhang Y., Tan B., Karlicky M., et al. // Astrophys. J. 799 (1). 30. 2015. 

Zhdanov A.A., Charikov Y.E. // Soviet Astronomy Letters. V.11.  88-90. 1985.

Zhou G.P., Zhang J., and Wang J.X. // Astrophys. J.823 (1). L19. 2016.

Zimovets I.V. et al. // Ge\& Ae. V. 62. I.4. P. 356-374. 2022. 

Zimovets I.V., Sharykin I.N., Kaltman T.I., et al. // Ge\& Ae. V.63. I.5. P.513-526. 2023

\end{document}